\documentclass{article}

\def\a{\alpha}
\def\b{\beta}
\def\ga{\gamma}
\def\de{\delta}   

\def\phi{\varphi}

\def\s{\sigma}
\def\z{\zeta}
\def\om{\omega}

\def\C{{\bf C}}

\def\h{{h}}

\def\Kb{{\bf K}}

\def\Mb{{\bf M}}

\def\R{{\bf R}}
\def\S{{\cal S}}
\def\T{{\rm T}}

\def\toro{{\bf T}}

\def\Om{\Omega}

\def\Cl{{\it C}\ell}

\def\pa{\partial}

\def\d{{\rm d}}       
\def\w{\wedge}

\def\ox{\otimes}
\def\o+{\oplus}

\def\grad{\nabla}     
\def\ss{\subset}

\def\<{\langle}
\def\>{\rangle}

\def\interno{\hskip 2pt \vbox{\hbox{\vbox to .18
truecm{\vfill\hbox to .25 truecm
{\hfill\hfill}\vfill}\vrule}\hrule}\hskip 2 pt}

\def\({\left(}
\def\){\right)}
\def\[{\left[}
\def\]{\right]}
\def\=#1{\bar #1}
\def\.#1{\dot #1}
\def\^#1{\widehat #1}
\def\"#1{\ddot #1}

\def\mapright#1{\smash{\mathop{\longrightarrow}\limits^{#1}}}

\def\en#1{\eqno(\sn.#1)}

\begin{document}

\title{{\bf Quaternionic integrable systems}}

\author{Giuseppe Gaeta \\ 
Dipartimento di Fisica, Universit\`a di Roma, \\
I--00185 Roma (Italy) \\ e-mail: {\it gaeta@roma1.infn.it} \\
~ \\
Paola Morando \\ 
Dipartimento di Matematica, Politecnico di Torino, \\
I--10129 Torino (Italy) \\ e-mail:  {\it morando@polito.it}}

\maketitle

{\bf Summary.} {Standard (Arnold-Liouville) integrable systems are intimately related to complex rotations. One can define a generalization of these, sharing many of their properties, where complex rotations are replaced by quaternionic ones. Actually this extension is not limited to the integrable case: one can define a generalization of Hamilton dynamics based on hyperKahler structures.}

\section{Introduction. Standard integrable systems}
\def\sn{1}

Consider an hamiltonian systems in $n$ degrees of freedom with compact energy manifolds, integrable in the Arnold-Liouville sense. 
By definition, these can be isomorphically mapped to an oscillators system, i.e. (no sum on $k=1,...,n$)
$$ {\dot p}_k \ = \ - \, \om_k \, q_k \ \ , \ \ {\dot q}_k \ = \ \om_k \, p_k \ ; \en{1} $$
hence can be mapped to a system manifestly equivariant under the standard torus action in $\R^{2n}$. Obviously, as the map mentioned above is a diffeomorphism, the system had to admit a torus action also in the original coordinates, and this is indeed guaranteed by the familiar condition of having $n$ integrals of motion in involution. 

In other words, Arnold-Liouville integrable systems with $n$ degrees of freedom are intimately related to symmetry under the group $U(1) \ox ... \ox U(1) = \toro^n$.
 
As well known, one can pass to action-angle coordinates $(I,\phi)$, defined via $p_k = \sqrt{I_k} \cos (\phi_k )$, $q_k = \sqrt{I_k} \sin (\phi_k)$; in these the evolution equations (1.1) read simply ${\dot I}_k = 0$, ${\dot \phi}_k = \om_k (I)$, and the $U(1) \ox ... \ox U(1) = [ U(1)]^{\ox n}$ symmetry is again immediately apparent.
 
We can equivalently consider, instead of the action-angle coordinates $(I,\phi)$, complex coordinates $z_k = \sqrt{I_k} e^{i \phi_k} = p_k + i q_k $; each of them evolves as 
$$ {\dot z}_k \ = \ i \, \om_k \, z_k \ , \en{2} $$ 
which of course has solution  $z_k (t) = e^{i \omega_k t} z_k (0)$.

Thus, the time evolution of an integrable system is given by a complex rotation -- with constant speed for given initial conditions -- in each $\C^1$ subspace (the frequency $\om_k$ depends in general on all  the $|z_k|^2 = I_k$, $k=1,...,n$). 

It is natural to expect that not much would change if instead of a {\it complex} rotation we had a {\it quaternionic} one. We show here that indeed this is the case. 

It is more surprising that one can develop a coherent theory of dynamical systems which are in a way a quaternionic generalization of standard Hamiltonian systems; these are more precisely related to {\it hyperkahler structures}, and thus will be therefore called {\it hyperhamiltonian}. 

Hyperkahler structures are briefly introduced e.g. in \cite{Ati,Cal}, see \cite{AtH,Swa} for a more complete discussion; they are the subject of increasing interest of mathematicians and physicists alike, see \cite{quat} for an updated overview (for the reasons of physical interest, see \cite{CFG,Hit}, and \cite{Zuc}; for connections with integrable systems, see e.g. \cite{Dan}). The construction of hyperhamiltonian dynamics is discussed in detail in \cite{GM}; see also \cite{MT} for the aspects related to quaternionic analysis \cite{Sud}.

\section{Generalizations: Clifford dynamics}
\def\sn{2}

Let us look at the relation between the $[U(1)]^{\ox n}$ symmetry and integrability in an even more explicit way: with $z_k = p_k + i q_k$, equation (1.2) is the familiar $n$-oscillator dynamics (1.1). With $\xi_k \in \R^2$ the two-dimensional vector $\xi_k = (p_k , q_k )$ and $J$ the standard two-dimensional symplectic matrix, this is rewritten as 
$$ {\dot \xi}_k \ = \ \om_k (|\xi_1|, ... , |\xi_n|) \ J \ \xi_k \ . \en{1} $$
 
In this notation, the integration of (1.2) goes simply through the fact that $|\xi_k|$ are constant (due to $J^T = - J$) and $J^2 = -I$, so that $\exp [\om_k J ] = \cos (\om_k t) + i \sin (\om_k t)$. Hence the solution of (\sn.1), which is $ \xi_k (t) = \exp[ \om_k J t] \xi_k (0)$,  reads
$$ \xi_k (t) \ = \ \[ \cos (\om_k t) \, I \ + \ \sin (\om_k t) \, J \] \ \xi_k (0) \ . \en{2} $$

Once we have looked at oscillator dynamics in such an elementary way, it is easy to find a direct (but non entirely trivial) generalization. 
 
Consider indeed $m$-dimensional vectors $\xi_k \in \R^m$ (we write $\rho_k = |\xi_k|^2$) and evolution equations of the form
$$ {\dot \xi}_k \ = \ \sum_{\a=1}^p \ \nu_{k\a} K_\a \xi \en{3} $$
where $\nu_{k \a} = \nu_{k \a} (\rho_1 , ... , \rho_p)$ are smooth functions, and $K_\a = K_\a (\rho_1 , ... , \rho_p)$ (with $\a = 1,...,p$) are $m$-dimensional matrices satisfying $K_\a^T = - K_\a$ and 
$$ \{ K_\a , K_\b \} \ = \ - \, 2 \, \delta_{\a \b } \, I \ . \en{4} $$
We will, for later reference, denote as $\Kb$ the Lie algebra spanned by the $K_\a$. Needless to say, (\sn.4) states that $\Kb$ is a {\it Clifford algebra}, with fundamental quadratic form $- I$; see e.g. \cite{Kir} or \cite{Hus}, chapter 11. 
 
It is immediate to check that under (\sn.3) the $\rho_k$ are conserved, due to $K_\a^T = - K_\a$; thus we can consider the $\nu_{k \a}$ and the $K_\a$ as constant on each trajectory of the system. 
 
The solution to (\sn.3) is of course $\xi_k (t) = \exp \[ \nu_{k\a} K_\a t \] \xi_k (0) $. When we evaluate the exponential, we have to take into account that, due to (\sn.4),
$$ \sum_{\a,\b = 1}^p \ \nu_{k\a} K_\a \nu_{k\b} K_\b \ = \ - \, \sum_{\a,\b = 1}^p \ \nu_{k \a} \nu_{k \b} \, \delta_{\a \b} \ . \en{5} $$
Hence, introducing the notation 
$$ \om_k \ = \ \( \sum_{\a=1}^p \nu_{k \a}^2 \)^{1/2} \ ; \ A_k :=
\sum_{\a=1}^p \ {\nu_{k\a} \over \om_k} \ K_\a \ , \en{6} $$
so that (\sn.3) is also rewritten as ${\dot \xi}_k = A_k \xi_k$, we have at once that 
$$ \xi_k (t) \ = \ \[ \cos (\om_k t) \, I \ + \ \sin (\om_k t) \, A_k \] \ \xi_k (0) \ . \en{7} $$
 
Thus, to any matrix algebra satisfying (\sn.4) we can associate a generalization of oscillator dynamics, integrable (or, to avoid any question like ``what is integrability?'', explicitely and elementarily solvable) by construction. We will also refer to these as {\bf Clifford} integrable systems.
 
As noted above, the $\xi_k (t)$ have constant norm under Clifford  dynamics; thus each $\xi_k \in \R^m$ evolves on the surface of a sphere $S^{m-1} \ss \R^m$ of constant radius $\rho_k = |\xi_k (0)|$, the $\rho_k$ can thus be considered as constant for any given initial data $\{ \xi_1 (0) , ... , \xi_p (0) \}$, and play the role of parameters.
 
We will now discuss the motion of a single vector $\xi_k \in \R^m$; we will assume $\rho_k \not= 0$ (or the motion is trivial), and write $\xi , \rho , A , ...$ for $\xi_k , \rho_k , A_k ,...$, for ease of notation. Also, as $\rho \not= 0$ is constant in time, we can assume $\rho = 1$ with no loss of generality, and the flow will take place on the unit sphere $S^{m-1} \ss \R^m$.

Consider $\xi \in S^{m-1} \ss \R^m$; the vectors $\{ K_1 \xi , ... , K_p \xi \}$ (spanning a space $\Kb (\xi)$), and hence the vector $A \xi$  belong to the tangent space $\T_\xi S^{m-1}$; in general, however, $\Kb (\xi)$ is a proper subspace of $\T_\xi S^{m-1}$ (i.e., in general $p < m+1$). This means that not all the directions of motion on $S^{m-1}$ are allowed, which is surely a limiting condition, and a too severe one to consider this a generalization of Hamilton dynamics.

There are however some cases in which $\Kb (\xi ) = \T_\xi S^{m-1}$ (i.e. $p=m-1$); these correspond to the existence of a Clifford algebra of dimension $m-1$ acting in $\R^m$, and of course a necessary (but not sufficient) condition for this to happen is that the sphere $S^{m-1}$ is parallelizable; this happens only for $S^1 , S^3, S^7$ \cite{Hus,Kir}. 

Indeed, the case $m=2$ corresponds to standard Hamilton dynamics (recall we are dealing with a single ``basic block'', i.e. in this case a minimal symplectic subspace); in the case $m=4$ we are dealing with $S^3 \ss \R^4$ and with the Clifford algebra ${\Cl} (2)$, well known to be isomorphic to the quaternion algebra ${\bf H}$, and in the following we will concentrate on this case. 
The case $m=8$ has not been explored yet as for the corresponding extension of Hamilton dynamics; note however that the corresponding Clifford algebra is ${\Cl} (3) \approx {\bf H} \oplus {\bf H}$ \cite{Hus,Kir}, and does not parallelize $S^7$. 
Thus the quaternionic case is the {\it only} full extension of standard Hamilton dynamics along the lines considered here.

\section{Quaternionic integrable systems}
\def\sn{3}

The simplest nontrivial Clifford algebra is just the quaternion one, i.e. ${\Cl} (2) = su(2)$; and here we want indeed to consider the quaternionic case. As it happens, this has several very special properties, shared with ${\Cl} (1) = U(1)$.

As we want to set our systems in the form (2.3), i.e. in $\R^4$ rather than in ${\bf C}^2$ or in ${\bf H}^1$, we should give a representation of the quaternionic imaginary units $i,j,k$ over $\R$; this is e.g. provided by the matrices
$$ \begin{array}{ll}
K_1 = \pmatrix{0&1&0&0\cr -1&0&0&0\cr 0&0&0&1\cr 0&0&-1&0\cr}
\ & , \ 
K_2 = \pmatrix{0&0&0&1\cr 0&0&1&0\cr 0&-1&0&0\cr -1&0&0&0\cr}
\ , \\ 
K_3 = \pmatrix{0&0&1&0\cr 0&0&0&-1\cr -1&0&0&0\cr 0&1&0&0\cr} \ , & \end{array} \en{1} $$
which satisfy -- as immediate to check -- the quaternion algebra
$$ K_\a K_\b \ = \ \varepsilon_{\a \b \ga} \, K_\ga \ - \ \de_{\a \b} \, I \en{2} $$

Note, for later use, that to these are associated the symplectic forms $\om_\a = (1/2) (K_\a)_{ij} \d x^i \w \d x^j$, given explicitely by 
$\om_1 = \d x^1 \w \d x^2  +  \d x^3 \w \d x^4$, $\om_2  = \d x^1 \w \d x^3 +  \d x^4 \w \d x^2$, and $\om_3 = \d x^1 \w \d x^4  +  \d x^2 \w \d x^3$; with $\Om$ the standard volume form in $\R^4$ we have $\om_\a \w \om_\a = 2 \Om$ (no sum on $\a$). 

In this case we rewrite (2.3) as 
$$ {\dot \xi} \ = \ \sum_{\a=1}^3 \ \nu_{\a} (\rho) \ K_\a \, \xi \ . \en{3} $$
and the general solution is immediately recovered from (2.7). This is 
$$ \xi (t) \ = \ \[ \cos (\om t) \, I \ + \ \sin (\om t) \, K \] \, \xi (0) \en{4} $$
and thus, for $\xi (0) \not= 0$, describes great circles $S^1$ on the sphere $S^3$ of radius $\sqrt{\rho} = | \xi (0) |$. This also means that this dynamics realizes the {\it Hopf fibration} $S^1 \to S^3 \to S^2$.

Note that the $K_\a$ are now constant matrices, and the dependence on $\xi$ (actually, on $\rho = |\xi|^2$) is only through the coefficients $\nu_\a$. This just means, of course, that we have chosen a basis $\{ K_1 , K_2 , K_3 \}$ in $\Kb \approx su(2)$.

We denote by $F$ the algebra of smooth real functions $f(\rho)$; this is of course the algebra of smooth functions $f: \R^4 \to \R$ which are constants on spheres $S^3$, i.e. invariant under rotations in $so(4) \approx su(2) \ox su(2)$.

It is quite interesting to note that we can rewrite (\sn.3) in a slightly different form. Introduce three functions $\h^\a : \R^4 \to \R$, with $\h^\a = h^\a (\rho) \in F$; now $\grad \h^\a = f^\a (\rho) \, \xi$, and we can hence rewrite (\sn.3) as
$$ {\dot \xi} \ = \ \sum_{\a=1}^3 \, K_\a \, \grad \h^\a \ , \en{5} $$
which makes completely transparent the relation with the hamiltonian case. 

Indeed, the flow $X$ described by (\sn.5) can be seen as the superposition of three hamiltonian flows $X_\a$, each of them defined by the hamiltonian $\h^\a$ with the symplectic structure $\om_\a$ associated to $K_\a$, see above. Obviously the $X_\a$ do not commute, but however they generate a module over $F$.

\section{Hyperhamiltonian dynamics}
\def\sn{4}

The equation (3.5) is the starting point for the extension of this setting to the non-integrable case. This can be defined on an arbitrary riemannian manifold $(M,g)$ of dimension $4n$ equipped with a {\it hyperkahler structure}. We will briefly recall how these are defined, and then define the associated hyperhamiltonian dynamics. We will work locally, i.e. on a single chart of $M$.

Let $(M,g)$ be a riemannian manidfold of (real) dimension $m = 4n$. 
Assume this is equipped with three complex structures $Y_\a$ ($\a = 1,2,3$), i.e. three (1,1) tensor fields such that $Y_\a^2 = - I$, satisfying the quaternionic relations
$$ Y_\a Y_\b \ = \ \epsilon_{\a \b \ga} Y_\ga \eqno(4.1) $$
(here and below $\epsilon$ is the completely antisymmetric Levi-Civita tensor). 

Assume moreover that $(M,g)$ is Kahler with respect to each of the $Y_\a$; in this case we say that $(M,g;Y_\a)$, or $M$ for short, is a {\it hyperkahler manifold}. 

We recall that $(M,g)$ is Kahler with respect to $Y_\a$ means that the Kahler form $\om_\a$ is closed, $\d \om_\a = 0$. The Kahler form is defined by 
$$ \om_\a (v,w) \ := \ g (Y_\a v , w) \ .  \eqno(4.2) $$

We can thus associate to each complex structure a symplectic structure $\om_\a$ by means of the Kahler relation (4.2); in this sense a hyperkahler structure (manifold) can also be seen as a ``hypersymplectic'' structure (manifold).

Consider now an ordered triple of arbitrary smooth functions $\h^\a : M \to \R$; we associate to these a triple of vector fields by (no sum on $\a$)
$$ X_\a \ \interno \ \om_\a \ = \ \d \, \h^\a \eqno(4.3) $$
and define the {\it hyperhamiltonian vector field} $X$ on $M$ associated to the triple $\{ \h^\a \}$ as the sum of these, 
$$ X \ := \ \sum_{\a=1}^3 \ X_\a \ ; \eqno(4.4) $$
it is trivial to check that the $X_\a$, and therefore $X$, are uniquely defined. Needless to say, the $X_\a$ do not commute; the relation between the $Y_\a$ guarantee they generate a module.

Let us now consider a local chart on $M$ and local coordinates $\{ x^1 , ... , x^m \}$ on this; we write as usual $\pa_i := (\pa / \pa x^i )$. 
The riemannian metric $g$ will be represented in coordinates by a (0,2) tensor field $g_{ij} (x)$, the complex structure $Y$ by a (1,1) tensor field $Y^i_{~j} (x)$, and the symplectic form by a (0,2) tensor field $W_{ij} (x)$, i.e. $\om = (1/2) W_{ij} (x) \d x^i \w \d x^j$. From now on we will omit to write down the dependence on $x$, for ease of notation.

The Kahler relation (4.2) implies that $ W_{ij} = g_{ip} Y^p_{~j}$. 
The relation $X \interno \om = \d H$ means that $ W_{\ell i}^T  X^i = \pa_\ell H $; as $W$ is nondegenerate it has an inverse and we write $(W^T)^{-1} := K$. With this we can also rewrite the relation above as $X^i = K^{ij} \pa_j H$. 

The hyperhamiltonian vector field will thus be 
$$ X^i \ = \ \sum_\a \ X_\a^i \ = \ \sum_\a \ K^{ij}_\a \ \pa_j \h^\a \ . \eqno(4.5) $$ 

Note that we have $K_\a = - g^{-1} Y_\a^T$. Indeed, $K^T = W^{-1} = (g Y)^{-1}$, and it is easy to check that this is just $K^T = - Y g^{-1}$, due to $Y^2 = - I$. Recalling that $g^T = g$, and similarly for their inverses, we get the statement. If $Y = - Y^T$ (which is in general not true), then $K = g^{-1} Y$.

Note also that the $K_\a$ satisfy the quaternion algebra with the natural multiplication between (2,0)-type tensor fields; it is immediate to check that $K_\a g K_\b = \epsilon_{\a \b \ga} K_\ga - \de_{\a \b} \eta$, with $\eta = g^{-1}$ the contravariant metric tensor.

It should be stressed that hyperhamiltonian dynamics shares many properties with standard Hamilton dynamics; among these we would like to point out in particular the possibility of a variational formulation \cite{GM}, which is a special case of a general situation:  Poincar\'e vector field can be described as characteristic fields of a variational principle based on maximal degree forms \cite{GM2}.

\section{Example. The Pauli spin equation}
\def\sn{5}

The natural physical application of the extension of hamiltonian mechanics to the hyperkahler case concerns, of course, spin systems.

The non-relativistic evolution equation for particles with spin one-half is provided by the Pauli equation. 
In the simplest case of a particle with spin 1/2 and considering only the spin degrees of freedom (we do not discuss here the physical meaning of this setting), this is written as 
$$ {d \Psi \over d t} \ = \ i \, \kappa \ ( {\bf B \cdot S} ) \, \Psi \ . \eqno(\sn.1) $$
Here $\kappa = 4 \pi \mu / h$ is a dimensional constant which we set to one in the following, $\Psi$ is a two-components spinor,
$$ \Psi \ = \ \pmatrix{\psi_+ \cr \psi_- \cr} \ \ , \ \ \psi_\pm (t) \in \C \ \ , \ \ \| \Psi \|^2 = 1 \ \ , \eqno(\sn.2) $$
the real vector ${\bf B}$ is the magnetic field, with components ${\bf B} (t) = (B_x , B_y , B_z)$, and ${\bf S}$ is the vector spin operator with components the Pauli $\s$ matrices. 
The linear operator $\Mb := {\bf B \cdot S}$ is given by
$$ \Mb \ = \ \pmatrix{B_z & B_x - i B_y \cr B_x + i B_y & - B_z \cr} \ . \eqno(\sn.3) $$

We want now to rewrite (\sn.1) as an equation in $\R^4$ rather than in $\C^2$. In order to do so, we rewrite $\psi_\pm$ separating their real and imaginary part as $\psi_\pm = \chi_\pm + i \z_\pm$; substituting a $\C^1$ number by an $\R^2$ vector,
$$ \psi_\pm \ = \ \pmatrix{\chi_\pm \cr \z_\pm \cr} \ , \eqno(\sn.4) $$
the operator of multiplication by $i$ is represented in $\R^2$ by the standard symplectic matrix, and we can use this to write $i \Mb$ as a real four-dimensional matrix (which we do in block notation):
$$ J \ = \ \pmatrix{0&-1\cr1&0\cr} \ \ , \ \  
i \Mb \approx \pmatrix{B_z J & B_y I + B_x J
\cr - B_y I + B_x J & - B_z J \cr} \ . \eqno(\sn.5) $$

Finally, the $\R^4$ representation of the Pauli equation is given by
$$ {d \xi \over dt} \ = \ A \, \xi \ , \eqno(\sn.6) $$
where
$$ \xi \, = \, \pmatrix{\chi_+ \cr \z_+ \cr \chi_- \cr \z_- \cr} \ \ , \ \
A \ = \ \pmatrix{ 0 & - B_z & B_y & - B_x \cr B_z & 0 & B_x & B_y \cr
- B_y & - B_x & 0 & B_z \cr B_x & - B_y & - B_z & 0 \cr} \ \ .
\eqno(\sn.7) $$

Let us define the matrices $\^K_\a$ as (the $\R^4$ representation of) 
$\^K_1 \simeq i \s_2$, $\^K_2 = i \s_1$, and $\^K_3 = i \s_3$. It is immediate to check that these satisfy, by construction, the quaternionic relations.

We can rewrite $A$ in terms of the matrices $\^K_\a$ as
$$ A (t) \ = \ B_y (t) \, \^K_1 \, + \, B_x (t) \, \^K_2 \, + \, B_z (t) \, \^K_3 \ . \eqno(\sn.8) $$
We see immediately from (\sn.7) and (\sn.8) that the $\R^4$ representation of the Pauli equation does correspond to a hyperhamiltonian system, with
$$ \begin{array}{l}
\h^1 (\xi,t) = (1/2) B_y (t) \| \xi \|^2 \ , \\ 
\h^2 (\xi,t) = (1/2) B_x (t) \| \xi \|^2 \ , \\ 
\h^3 (\xi,t) = (1/2) B_z (t) \| \xi \|^2 \ . \end{array} \eqno(\sn.9)$$

When ${\bf B}$ does actually not depend on $t$, we have an integrable system. If ${\bf B}$ varies with $t$, we have explicitely time-dependent hamiltonians $\h^\a (|\xi|^2 ; t)$: the system is not integrable, but $|\xi|$ is still constant. 

This is coherent with -- and actually required by -- the physical meaning of $\xi$, which represents components of the spin wave function: its total square modulus $|\xi|^2 = \sum_\a |\xi^\a|^2$ must be constant, and equal to one by normalization.

It may be interesting, in the present context, to point out that the equation (\sn.1) was analyzed by Cari\~nena, Grabowski and Marmo in the context of nonlinear superposition principles \cite{CGM,ShW}. 

\section{Some final comments}

In this note we have presented a very streamlined version of our approach. We would now like to briefly present some final comments concerning relation to the complete picture. These will be rather rough due to lack of space, see \cite{GM} for further detail.
\medskip

{\bf (1).} First of all, it is natural to say that a general  hyperhamiltonian dynamical vector field $X$, see (4.3) and (4.4), on $(M,g)$ with coordinates $x^1, ... , x^{4n}$ is integrable if there is a diffeomorphism $\Phi : x \to \xi$ (where $\xi = \xi^1,...,\xi^{4n}$) such that if we pass to $(\rho_s ; \varphi_1^s , \varphi_2^s , \varphi_3^s)$ coordinates (with $s = 1,...,n$), the flow $X$ is described by (3.6) with $K$ and $\h$ depending only on $\rho$, i.e. is a ``quaternionic oscillator''. 
Note that to guarantee integrability of such a hyperhamiltonian system it suffices to have $n$ independent integrals of motion (the $\rho_1,...,\rho_n$), rather than $2n$ as it would be the case for hamiltonian systems.
\medskip

{\bf (2).} As very briefly mentioned above, hyperhamiltonian dynamics admits a variational formulation, albeit a non-standard one. This is based on a double fibration of the enlarged phase space $\^M := M \times \R$ (the $\R$ factor should be thought of as the time), i.e.
$ \^M \ \mapright{\pi} \^B \mapright{\tau} \R $
with $\^B = B \times \R$ a smooth manifold of dimension $4m-1$. 
The variational principle is expressed by choosing an arbitrary domain $C \ss \^B$ and requiring zero variation of the functional 
$$ I (\Phi) \ := \ \int_C \Phi^* (\theta ) $$
defined on sections $\Phi$ of the bundle $\^M \mapright{\pi} \^B$ under the action of vector fields wich are vertical for this bundle and vanish on $\pa C$. 
Here $\Phi^*$ is the pullback of $\Phi$, and $\theta \in \Lambda^{4n-1} (\^M)$ is a form defined by
$ \theta := \psi  +  (6 ns)  \sum_\a  \h^\a  \zeta_\a \w \d t$, 
where $s = \pm 1$ depending on orientation matters, the $(2n-2)$-forms $\zeta_\a$ are defined as $\zeta_\a = \om_\a \w ... \w \om_\a$ with no sum on $\a$ and, with $\d \s_\a = \om_\a$,  
$ \psi := \sum_{\a =1}^3  \s_\a \w \zeta_\a $. 
The forms $\psi$ and $\theta$ play respectively the role of the Poincar\'e form and Poincar\'e-Cartan invariant in hyperhamiltonian dynamics.
\medskip

{\bf (3).} It is quite obvious that we could operate a permutation of the $\om_\a$, or also act with a rotation on the three dimensional linear space ${\bf U}$  spanned by these, and -- provided a similar operation is performed on the $\h^\a$ -- leave the hyperhamiltonian flow $X$ invariant. 

This suggest that the triple of symplectic structures $\om_\a$ should not be seen as the central geometrical object, and indeed the natural invariant object characterizing the theory is the unit sphere $\S \simeq S^2 \ss U$.
This has a counterpart, via the metric $g$ and the Kahler relation, in the unit sphere in the linear space ${\bf O}$ spanned by the complex structures $Y_\a$; the space ${\bf O}$ is also said to be a quaternionic structure on $M$, and if the $Y_\a$ are unimodular then its unit sphere corresponds to unimodular complex structures in $(M,g)$ \cite{AlM}.

Note that the transformations preserving the structure (``canonical transformations'') will not be those preserving the triple $\om_\a$ (i.e. triholomorphic ones), but instead those mapping the sphere $\S \ss U$ into itself, i.e. such that $f^* : \S \to \S$ although in general with $f^* (\om_\a ) \not= \om_\a$.

This discussion shows at once -- at least to the reader familiar with that approach -- the relation of hyperhamiltonian systems, in particular integrable ones, with twistors theory.
 

\end{document}